\documentclass[12pt]{article}
\usepackage{latexsym,graphicx}
\usepackage{hfstyle}
\newtheorem{conjecture}{Conjecture}
\newcommand{\F}{{\cal F}}
\newcommand{\B}{{\cal B}}
\newcommand{\G}{{\cal G}}
\newcommand{\calS}{{\cal S}}
\newcommand{\f}{{\rm \bf f}}
\newcommand{\g}{{\rm \bf g}}
\newcommand{\h}{{\rm \bf h}}
\newcommand{\0}[1]{{{\bf 0}^{(#1)}}}
\newcommand{\1}[1]{{{\bf 1}^{(#1)}}}
\newcommand{\Z}{{\mathbf{Z}}}
\newcommand{\N}{{\mathbf{N}}}
\newcommand{\I}{{\mathrm{I}_n}}
\newcommand{\card}{\mathop{\rm{card}}}
\newtheorem{proposition}{Proposition}[section]
\begin{document}
\title{Sequences of preimages in elementary
cellular automata}
\author{Henryk Fuk\'s
      \oneaddress{
  Department of Mathematics,
   Brock University,
 St. Catharines, ON, Canada\\[1em]
         \email{hfuks@brocku.ca}
       }
   }

\Abstract{We search for regularities in the sequences of numbers of
preimages for elementary cellular automata. For 46 out of 88
``minimal'' rules, we find recognizable patterns, usually in the
form of second order recurrence equations with constant
coefficients. Introducing the concept of asymptotic emulation of
CA rules, we then show how the regularities in the  sequences of
preimage numbers can be used to find rules emulating identity. We
also show that the average density of nonzero sites after
arbitrary number of steps (starting from disordered configuration)
can be computed using the sequences of preimage numbers.
}
\maketitle

\section{Introduction}

One of the fundamental problems in the theory of cellular automata is
the problem of enumeration of preimages. Preimages for a given spatial
 sequence are defined as the set of blocks that are mapped
to that sequence by the automaton rule. Since the number of
preimages for the sequence provides information about the
probability distribution associated with the rule, it can be
useful for a variety of problems, like computations of spatial
measure entropy, identification of sequences with maximal
probability \cite{Jen89}, identification of the Garden of Eden
\cite{Voorhees96}, etc.

For one-step preimages, E. Jen \cite{Jen88} showed that the number
of preimages for arbitrary sequences satisfies a system of
recurrence relations with coefficients depending on the automaton
rule. No analytical results, however, are known for the number of
$n$-step preimages, i.e. the number of preimages under the rule
iterated $n$ times. In this paper, we will show that the
sequences of $n$-step preimage numbers in many cases follow
recognizable patterns, so the expression for the general term of
the sequence can be conjectured (and, in some simple cases,
proved). We will then present two possible applications of such
expressions, in finding asymptotical emulators of CA rules and
densities of nonzero sites after arbitrary number of time steps.

Let $\G=\{0,1,...N-1\}$ be called {\it a symbol set}, and let $\calS(\G)$
 be the set of all bisequences over $\G$, where by a bisequence we mean a
 function on  $\Z$ to $\G$. Set  $\calS(\G)$, which
 is a compact, totally disconnected, perfect, metric space,  will be
called {\it the configuration space}. Throughout the remainder of this
paper
 we shall  assume that $\G=\{0,1\}$, and  the configuration space
 $\calS(\G)=\{0,1\}^{\Z}$ will be simply denoted by $\calS$.

{\it A block of radius} $r$ is an ordered set $b_{-r} b_{-r+1}
\ldots b_r$, where $r\in \N$, $b_i \in \G$. Let $r\in \N$ and let
$\B_r$ denote the set of all blocks of radius $r$ over $\G$. The
number of elements of $\B_r$ (denoted by $\card \B_r$) equals
$2^{2r+1}$. The set of all
blocks of finite radius will be denoted by $\B=\bigcup_{r=0}^{\infty}\B_r$.

A mapping $f:\{0,1\}^{2r+1}\mapsto\{0,1\}$ will be called {\it a cellular
 automaton rule of radius $r$}. Alternatively, the function $f$ can be
 considered as a mapping of $\B_r$ into $\B_0=\G=\{0,1\}$. The set of all
mappings of radius $r$ will be denoted by $\F_r$, and the set of all possible
cellular automata mappings by $\F=\bigcup_{r=0}^{\infty}\F_r$.

Corresponding to $f$ (also called {\it a local mapping}) we define a
 {\it global mapping}  $F:S\to S$ such that
$
(F(s))_i=f(s_{i-r},\ldots,s_i,\ldots,s_{i+r})
$
 for any $s\in S$.
The {\it composition of two rules} $f,g\in \F$ can be now defined in terms
of
their corresponding global mappings $F$ and $G$ as $
(F\circ G)(s)=F(G(s)),
$
where $s \in S$. We note that if  $f \in \F_p$ and $g \in \F_q$,
 then $f \circ g \in \F_{p+q}$. For example, the composition of two
radius-1
mappings is a radius-2 mapping:
\begin{equation}
(f\circ g)(s_{-2},s_{-1},s_0,s_1,s_2)=
f(g(s_{-2},s_{-1},s_0),g(s_{-1},s_0,s_1),g(s_0,s_1,s_2)).
\end{equation} Multiple composition will be denoted by
\begin{equation}f^n=\underbrace{f \circ f
 \circ \cdots \circ f}_{\mbox{$n$ times}}.
\end{equation}

A {\it block evolution operator} corresponding to $f$ is a mapping
 $\f:\B \mapsto \B$ defined as follows. Let $r \geq p >0$, $a\in \B_r$,
$f\in \F_p$, and   let $b_i=f(a_{i-p},a_{i-p+1},
\ldots,a_{i+p})$ for $-r+p \leq i \leq r-p$. Then we define $\f(a)=b$,
 where  $b \in \B_{r-p}$.
 Note that if
$b \in B_1$ then $f(b)=\f(b)$.

In what follows we will consider the case of $\G=\{0,1\}$ and $r=1$ rules,
 i.e. {\it elementary cellular automata}. The set of radius-1 blocks
$\B_1$
has then $8$ elements, which will be denoted by
\begin{equation}\{\beta_i\}_{i=0}^{i=7}=\{000,001,010,011,100,101,110,111\},
\end{equation} so that the binary representation of the index $i$ defines the block
$\beta_i$. Given an elementary rule $f$, we will try now to find
the number of $n$-step preimages of such {\it basic blocks}
under the rule $f$.
\section{Sequences of Preimage Numbers}
The number of $n$-step preimages of the block $b$ under the rule $f$
is defined as the number of elements of the set $\f^{-n}(b)$. For
surjective rules, this number is always easily computed. As proved
in \cite{hedlund69}, under the surjective elementary rule every block has exactly
four preimages, so $\card [\f^{-n}(b)]=4^n$ for every block $b$.
For non-surjective rules, however, sequences of
n-step preimage numbers can be highly nontrivial, and no general method
for obtaining them without direct counting is known.

Using a simple preimage counting computer program, sequences $a_n
=\card \f^{-n}(b)$ can be constructed for a given rule $f$ and a
block $b$. For some elementary cellular automata rules and basic
blocks, these sequences appear to follow certain recognizable
patterns, while for other rules no pattern seems to emerge after
computation of the first $10$ terms (since the number of possible
blocks increases exponentially with the block length, it becomes
increasingly difficult go much beyond $n=10$ using direct enumeration).
\begin{table*}
\begin{center}
\begin{tabular}{||c|l||} \hline\hline
Block $b$ & Rule numbers of the rule $f$ \\ \hline \hline
001&    0 \\ \hline
010&    0,19,46,126,200 \\ \hline
011&    0,2,4,8,12,24,32,34 \\ \hline
100&    0 \\ \hline
101&    0,1,2,3,8,10,11,36,128,136,138 \\ \hline
110&    0,2,4,8,12,24,32,34 \\ \hline
111&    0,2,4,6,8,10,12,14,18,24,28,32,34,40,42,50,56,72,76 \\ \hline \hline
\end{tabular}
\end{center}
\caption{Blocks $b$ which have no preimages under some elementary
rules $f$.} \label{nopre}
\end{table*}
The simplest pattern to recognize is the constant sequence $a_n= const$.
Let us first consider the case when $\f^{-n}(\beta_i)$ is empty,
i.e., $a_n=0$ for every positive integer $n$.  In order to prove
that a given block $b \in \B_1$ has no preimage under a given rule
(i.e. $\card \f^{-n}(\beta_i)=0$). One just has to check that for
every block $c\in \B_2$ (among 32 possible) condition $\f(b) \neq
c$ is satisfied. We performed this check for all 88 minimal
elementary cellular automata rules and all basic blocks. Results
are presented in Table~\ref{nopre}.

Another type of the constant sequence is the case when the set
$\f^{-n}(\beta_i)$ has only one element regardless of $n$, i.e. $a_n=1$.
All such cases are shown in
Table \ref{table2}. Although this table was generated with the help of
a computer, it is not difficult to prove that $\card \f^{-n}(b)=1$
for a given basic block $b$.
As an example, consider elementary rule 77 (for this rule,
$\f^{-1}(0)=\{001, 100, 101, 111\}$). We claim that
\begin{proposition}
\label{prop77}
For the rule 77, both sets $\f^{-n}(000)$ and $\f^{-n}(111)$
have only one element for all positive integers $n$.
\end{proposition}
To see it, let us consider a block of ones $11 \ldots 1$ of radius $r$ ,
which will be denoted by  $\1{r}$ (similarly, block of zeros of radius
$r$ will be denoted by $\0{r}$). It is easy to verify that
$\f^{-1}\left(\1{r}\right)=\0{r+1}$. Indeed, if we assume that there exists
a block $a\in\B_{r+1}$ such that $\f(a)=\1{r}$, with at least one nonzero site, then block $a$ must include at least one of subblocks 001, 100,
 101 or 111. All these subblocks belong to $\f^{-1}(0)$, so $\f(a)$ cannot
be $\1{r+1}$. Therefore, for the rule 77 $\f^{-1}\left(\1{r}\right)=\0{r+1}$,
and similarly $\f^{-1}\left(\0{r}\right)=\1{r+1}$, what implies that
$\f^{-n}(000)$ and $\f^{-n}(111)$ are single-element sets for every
positive integer $n$.Similar proof can be constructed for other entries
 in Table \ref{table2}.
\begin{table*}
\begin{center}
\begin{tabular}{||c|l||}\hline
Block $b$  &  Rule numbers of the rule $f$\\ \hline \hline 000 &
77, 178 \\ \hline 001 & none \\ \hline 010 & 23,128,232 \\ \hline
011 & 128,160, 162, 130, 132 \\ \hline 100 & none \\ \hline 101 &
23, 32, 44, 130, 232, 33 \\ \hline 110 & 128, 130, 132, 160, 162,
\\ \hline 111 & 77, 128, 130, 132, 134, 146, 160, 162, 178 \\
\hline
\end{tabular}
\end{center}
\caption{Basic blocks $b$ and rules $f$ for which $\card [\f^{-n}(b)]=1$
for every positive integer $n$. }
\label{table2}
\end{table*}

Cases with $a_n=\mathrm{const}>1$ are not numerous. We found only seven
of them in all ``minimal'' elementary rules, with the largest possible
constant $a_n$ equal to 5.
These cases can be  summarized in the following
conjecture:
\typeout{Check an=1 case again}
\begin{conjecture}
The only minimal elementary cellular automata rules and the only
basic blocks for which
the sequence of preimage numbers
$a_n=\card [\f^{-n}(\beta_i)]$ is constant (i.e. $n$-independent) and
$a_n>1$ are:
\begin{itemize}
\item $\card [\f_{128}^{-n}(001)]=\card [\f_{128}^{-n}(100)]=2$
\item $\card \f_{32}^{-n}(001)=\card [\f_{32}^{-n}(100)]=
\card [\f_{58}^{-n}(000)]=3$
\item $\card [\f_{32}^{-n}(010)]=4$
\item $\card [\f_{50}^{-n}(000)]=5$
\end{itemize}
(All the above expressions hold for any positive integer $n$)
\end{conjecture}

The sequence $a_n$ can be, of course, much more complicated that
$a_n=\mathrm{const}$. After experimenting with various
possibilities, we found that in many cases $a_n$ appears to
satisfy a second order difference equation with constant
coefficients \begin{equation}a_{n+2}=c_1 a_{n+1} + c_2 a_n +c_3.
\label{difeq} \end{equation} To check whether it is plausible, we
performed the following test. Using first 5 terms of $a_n$
 (obtained using the preimage counting program) we can
solve the system of 3 linear equations for $c_1, c_2, c_3$:
\begin{eqnarray}
\label{seqtest}
a_{3}&=&c_1 a_{2} + c_2 a_1 +c_3, \\ \nonumber
a_{4}&=&c_1 a_{3} + c_2 a_2 +c_3, \\
a_{5}&=&c_1 a_{4} + c_2 a_3 +c_3. \nonumber
\end{eqnarray}
The solution $c_1, c_2, c_3$ can be now used to generate the next five
terms of the sequence $a_6 \ldots a_{10}$. If they agree with the
experimental values of $a_6 \ldots a_{10}$, we can conjecture that
the sequence $a_n$ is a solution of the difference equation~(\ref{difeq}).

As an example, let us consider the rule 172 and the block $101$. This
block has 4 preimages under $\f_{172}$, 12 preimages under $\f^2_{172}$,
40 preimages under $\f^3_{172}$ etc. The first few terms of
$a_n=\card[\f^{-n}_{172}(101)]$ obtained using the preimage counting
program are
\begin{equation}a_n=\{2,6,20,64,208,672,2176,7040,22784 \ldots \}
\end{equation} Solving (\ref{seqtest}) we obtain $c_1=2$, $c_2=4$, $c_3=0$, i.e.
\begin{equation}a_{n+2}=2a_{n+1} + 4a_n.
\end{equation} Although this difference equation was obtained using $a_1 \ldots a_5$ only,
it is easy to check that it is satisfied for all 10 term listed above.
Its solution is
\begin{equation}a_n=\frac{(1+\sqrt{5})^{n+2}-(1-\sqrt{5})^{n+2}}{8 \sqrt{5}}. \label{r172101}
\end{equation}
The same procedure can be applied to other elementary rules, and for
many of them expressions similar to (\ref{r172101}) can be found.
A table in the Appendix  shows all such cases.
They are presented as a set of 8 expressions, each
representing $a_n=\card [\f^n(\beta_i)]$ for all 8 basic blocks $\beta_i$,
$i=1 \ldots 7$. Only rules for which we were able to conjecture
all 8 expressions are shown, including cases when $a_n=\mathrm{const}$.
Surjective rules ( i.e. 15, 30, 45, 51, 60, 90, 105, 106, 150, 154, 170 and 204)  are omitted, since for
them we always have $a_n=4^n$.

\section{Asymptotic Emulation in Cellular Automata}
We say (after \cite{Rogers94}) that $f$ {\it emulates} $g$ in $k$ iterations
($k\geq0$) or $f$ is {\it a $k$th level emulator of $g$} if
\begin{equation}f\circ f^k=g \circ f^k. \label{emdef} \end{equation} If a cellular automaton $f$ emulates $g$ then after $k$ time steps we can
replace the rule $f$ by $g$ and we will obtain the same result as if we
had kept
rule $f$.  For example, many elementary ($r=1$) rules
emulate the identity rule. As proved in \cite{Rogers94}, these rules are
 0, 4, 8, 12, 36, 72, 76, 200, and 204 (only minimal representatives are
 listed here), and the level of emulation is always 0, 1, or 2.
Spatiotemporal patterns generated by these rules after a few time steps
 become identical with the pattern generated by identity rule (vertical
strips), as shown in Figures 1a and 1b.
 \begin{figure}
\begin{center}
\begin{tabular}{ll}
 a) \includegraphics[scale=0.8]{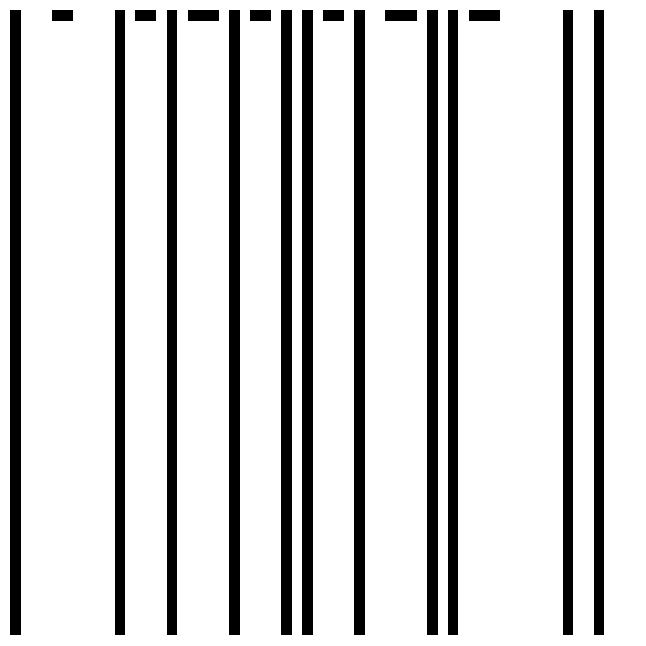}&
 b) \includegraphics[scale=0.8]{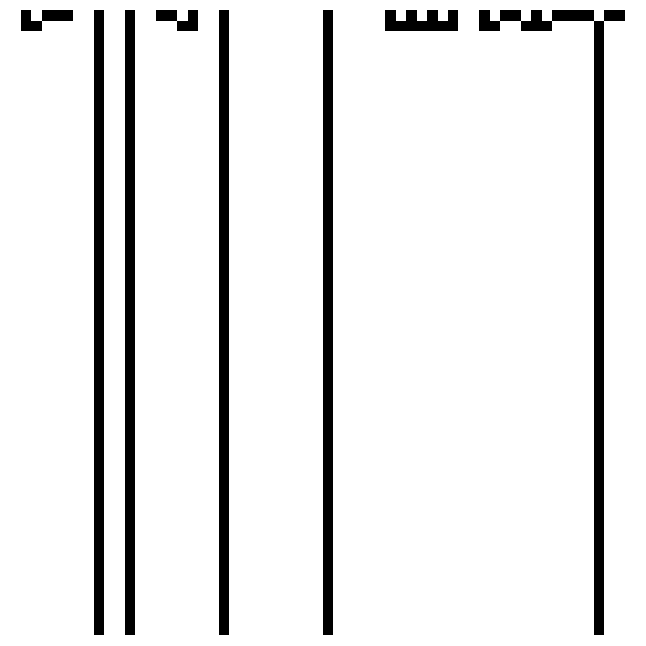} \\
 c) \includegraphics[scale=0.8]{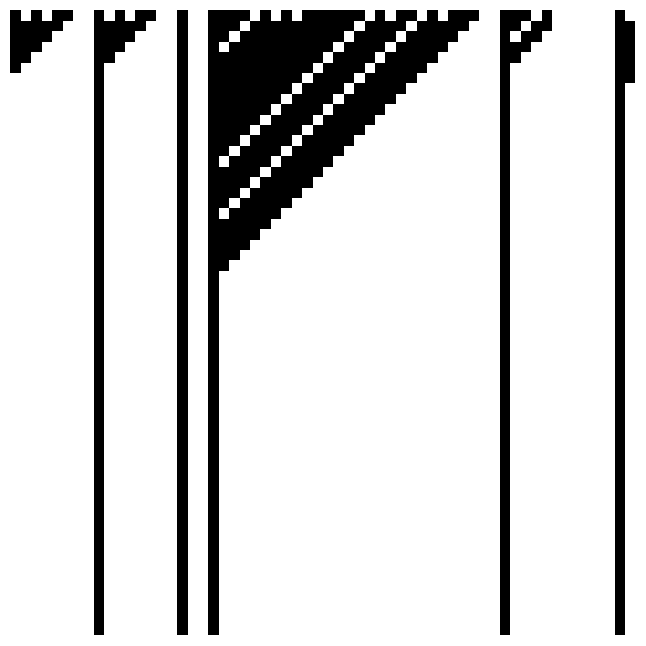}&
 d) \includegraphics[scale=0.8]{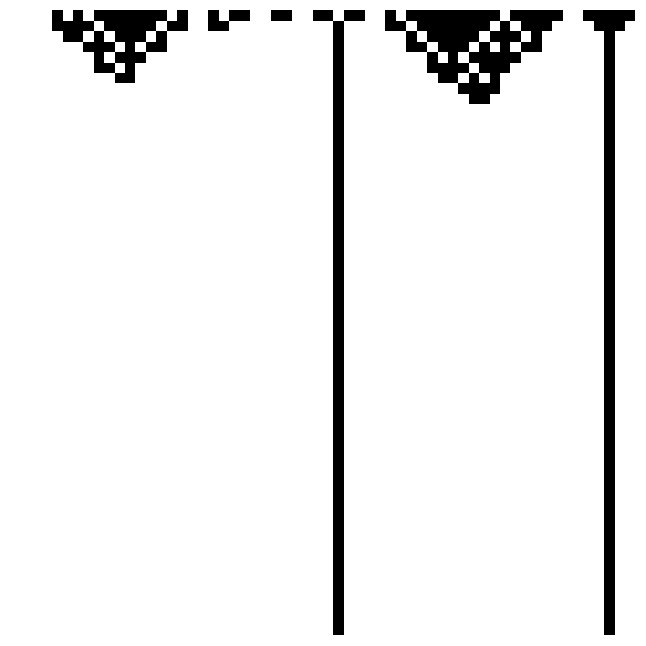}
\end{tabular}
\end{center}
\caption{Examples of cellular automata rules emulating identity: a) rule 4,
first-level emulation, b) rule 36, second-level emulation, c) rule 172,
asymptotic emulation, d) rule 164, asymptotic emulation.}
\end{figure}
 Visual examination of patterns generated by elementary cellular automata
reveals that not only rules mentioned earlier produce patterns resembling
rule 204 (identity rule). Among 88 ``minimal'' representatives of elementary
 rules there are 16 other ``identity-like'' mappings, namely 13, 32, 40, 44,
 77, 78, 104, 128, 132, 136, 140, 160, 164, 168, 172, and 232.
Typical patterns produced by these mappings are shown in Figures 1c and
1d. These
patterns eventually become vertical strips, but time required to achieve
such a state may be quite long. None of them, of course, emulates identity
in the sense of definition \ref{emdef}. We could say, however, that these
 rules simulate identity ``approximately'', and that this approximation is
getting better and better with increasing number of time steps.
Quantitative description of this phenomenon is possible if we introduce a
 distance between rules.
For $f\in\F_p$ and $b\in\B_q$, where $q>p$,  we define
$f(b)=f(b_{-r},\ldots,b_i,\ldots,b_r)$. Metric in $\F$ can be constructed
 as follows:
\begin{proposition}
Let $f\in\F_m$, $g\in\F_n$, and $k=\max\{m,n\}$. A function
$d:\F\times\F\mapsto[0,1]$ defined~by
\begin{equation}d(f,g)=2^{-2k-1}\sum_{b\in\B_k}\left|f(b)-g(b)\right|
\label{dist}
\end{equation}  is a metric in $\F$.
\end{proposition}
Obviously, $d(f,g)\geq 0$ and $d(f,g)=0 \Leftrightarrow f=g$. Triangle
inequality holds too since $|x+y|\leq|x|+|y|$ for all $x,y\in\{0,1\}$.

A cellular automaton rule $f$ {\it asymptotically emulates
 rule $g$} if
\begin{equation}\lim_{n\to \infty}d(f^{n+1}, g\circ f^n)=0.
\end{equation} Clearly, if $f$ is a $k$th level emulator of $g$  then $f$ emulates $g$
asymptotically. We may think about asymptotic emulation as $\infty$th level
emulation.

Let us now consider two rules $f,g\in\F$. Their {\it sum modulo} 2
will be defined as $(f\oplus g)(b)=f(b) + g(b) \mathrm{mod\ } 2
=|f(b)-g(b)|$ for any $b \in \B$. Note that $(f\oplus g)(b)=0$ if
$f(b)=g(b)$ and $(f\oplus g)(b)=1$ if $f(b)\neq g(b)$.

\begin{proposition}
\label{main}
Let $f , g \in \F_1$ and $h=f \oplus g$. Let $A_0=\h^{-1}(1)$, and let $A_n=
\f^{-n}(A_0)$. Then
\begin{equation}d(f^{n+1}, g\circ f^n)=\frac{\card A_n  }{2^{2n+3}}.
\end{equation} \end{proposition}
{\it Proof.} Mapping $f^{n+1}$ is a rule of radius $n+2$, therefore using the
definition of the distance (\ref{dist}) and properties of block evolution
 function we have
\begin{equation}d(f^{n+1}, g\circ f^n)= 2^{-2n-3}\sum_{b\in\B_{n+1}}\left|\f^{n+1}(b)-\g\circ
\f^n(b)\right|,
\label{eq6}
\end{equation} or $d(f^{n+1}, g\circ f^n)= 2^{-2n-3}c_n$, where $c_n$ is a number of blocks
$b \in \B_{n+2}$ such that $\f^{n+1}(b) \neq \g\circ \f^n(b)$.
Similarly, the
set $A_0$ is a set of all blocks $b\in \B_1$ such that $\f(b) \neq \g(b)$.
Let us now consider a block $a\in B_{n+1}$ such that
$\f^{n+1}(a) \neq \g\circ \f^n(a)$.
The last relation can be written as $\f\left(\f^n(a)\right)
 \neq \g\left(\f^n(a)\right)$, and this is possible iff $\f^n(a)
 \in A_0$, which is equivalent to $a \in \f^{-n}(A_0)$. This proves that
block $a\in \B_{n+1}$ satisfies  $\f^{n+1}(a) \neq \g\circ \f^n(a)$ iff $a \in A_n$, so finally $c_n= \card A_n$. $\Box$

Proposition \ref{main} can be  useful in finding asymptotical emulators.
As an example, consider the case of rule 77 discussed earlier, where we have
\begin{equation}A_0=(\f_{77}\oplus \f_{204})^{-1}(1)
=\{000,111\},
\end{equation} We already proved (in Proposition \ref{prop77}) that
both $\f_{77}^{-n}(000)$ and $\f_{77}^{-n}(111)$
have only one element for all $n$. Note that
\begin{equation}\card[ \f_{77}^{-n}\{000,111\}]=\card [f_{77}^{-n}(000)] +
\card[\f_{77}^{-n}(111)]=1+1=2,
\end{equation} since the preimage of the union of two set is always the union of the
preimages of the sets. This leads to the conclusion that
\begin{equation}d(f_{77}^{n+1}, f_{204}\circ f_{77}^n)=\frac{2}{2^{2n+3}}=2^{-2n-2}.
\end{equation} Of course, the above distance goes to zero with $n$, therefore rule
77 asymptotically emulates the identity (rule 204). Almost identical
reasoning
can be presented for rules 128 and 132, both of which asymptotically
emulate identity and
\begin{eqnarray}
d(f_{128}^{n+1}, f_{204}\circ f_{128}^n)& = & 3 \cdot 2^{-2n-3} \nonumber \\
 d(f_{132}^{n+1}, f_{204}\circ f_{132}^n)& = & 2^{-2n-2}.
\end{eqnarray}

Slightly different analysis can be performed for rule 32. Here,
from Table \ref{nopre}, we read that $\card [\f_{32}^{-n}(101)]=1$. Since
\begin{eqnarray}
(\f_{32}\oplus \f_{0})^{-1}(1)
=101, \nonumber
\end{eqnarray}
we conclude that $d(f_{32}^{n+1}, f_{0}\circ f_{32}^n)=2^{-2n-3}$, and
therefore rule 32 emulates the zero rule asymptotically.
It also emulates the identity rule asymptotically, as a consequence of the
following general property:
\begin{proposition}
If $f \in \F$  emulates the zero rule asymptotically,
then it also emulates the identity rule asymptotically.
\end{proposition}
Using the triangle inequality, we have
\begin{equation}0\leq d(f^{n+1}, f_{204}\circ f^n) \leq
d(f^{n+1}, f_{0}\circ f^n)+
d( f_{0}\circ f^n, f_{204}\circ f^n).
\end{equation} Since $f_{0}\circ f^n=f_0$ and $f_{204}\circ f^n=f^n$, we obtain
\begin{equation}d( f_{0}\circ f^n, f_{204}\circ f^n)=d(f_0,f^n)=d(f^{n},
f_0 \circ f^{n-1}).
\end{equation} The above equation, and the fact that $f$ asymptotically emulates $f_0$,
implies
 \begin{eqnarray*}
\lim_{n\to \infty}{\left(d(f^{n+1}, f_{0}\circ f^n)+
d( f_{0}\circ f^n, f_{204}\circ f^n)\right)}& = & \\
\lim_{n\to \infty}d(f^{n+1}, f_{0}\circ f^n)+
\lim_{n\to \infty}d(f^{n},f_0 \circ f^{n-1})& = & 0,
\end{eqnarray*}
so finally $\lim_{n\to \infty} d(f^{n+1}, f_{204}\circ f^n)=0$, as required
 for $f$ to emulate identity asymptotically. $\Box$

Of course, we could directly use expressions from the Appendix
 and find that
\begin{equation}d(f_{32}^{n+1}, f_{204}\circ f_{32}^n)  = \frac{5}{2^{2n+3}} .
\end{equation}

For other identity-like rules mentioned at the beginning of this section,
mechanism of emulation is not as simple as in previous cases. Nevertheless,
experimental evidence suggest the following
conjecture:
\begin{conjecture}
Among the $88$ ``minimal'' elementary cellular automata rules, only rules
$14,40,44,78,104,136,140,160,164$ and $172$ asymptotically emulate
the identity rule.
\end{conjecture}
Postulated expressions for the distance $d(f^{n+1}, f_{204}\circ f^n)$
are shown in Table \ref{idemul}. For completeness, rules for which the
proof
is known (i.e. 32, 77, 128,
and 132) are included as well.
\begin{table*}
\begin{tabular}{||c|l||c|l||}\hline
$f$ & $d_n=d(f^{n+1}, f_{204}\circ f^n)$ & $f$ & $d_n=d(f^{n+1},f_{204}\circ f^n)$ \\ \hline \hline
13 & $7\cdot 2^{-n-4} $ & 132 & $2^{-2n-2}$ \\ \hline
32 & $5\cdot 2^{-2n-3}$ & 136 & $2^{-n-2}$ \\ \hline
40 & $2^{-n-1}        $ & 140 & $2^{-n-3}$ \\ \hline
44 & $7\cdot 2^{-2n-3}$ & 160 & $3\cdot2^{-n-2} - 4^{-n-1}$ \\ \hline
77 & $2^{-2n-2}$         & 164 & $5\cdot 2^{-n-3} -4^{-n-1}$ \\ \hline
78 &  $4^{-1}$ if $n=1$  & 168 & $ 3^{n+1}\cdot 2^{-2n-3}$ \\
   &  $15 \cdot 2^{-n-6}$ if $n>1$   &     &         \\ \hline
104 & $163\cdot 2^{-2n-3}$ if $n>5$
 & 172 & $\frac{-(1-\sqrt{5})^{n+3} +
 (1+\sqrt{5})^{n+3}}{2^{2n+6}\sqrt{5}}$ \\ \hline
128 & $3\cdot 2^{-2n-3}$ & 232 & $2^{-2n-2}$ \\ \hline
\end{tabular}
\caption{Distance $d_n=d(f^{n+1}, f_{204}\circ f^n)$ for rules
asymptotically emulating identity.}
\label{idemul}
\end{table*}
\section{Density of nonzero sites}

The simplest statistical quantity characterizing a configuration is
the average fraction of sites with value~$1$ at time $t$, denoted by $c_t$.
The question we want to address now is as follows:
If we start from a disordered configuration with $c_0=0.5$ (i.e. equal
probability of 0 and 1), what is the density $c_t$ at a later time~$t$?
When $c_0=0.5$, a disordered configuration contains all 8 possible basic
blocks with equal probability. Applying a cellular automaton rule to this initial state
yields a configuration in which the fraction of sites with value 1 is
given by
\begin{equation}c_1=\frac{\card [\f^{-1}(1)] }{8},
\end{equation} or in other words, by the fraction of the eight possible basic blocks
which yield 1 according to the cellular automaton rule \cite{Wolfram94}.
Similarly, the density of ones after two time steps will be given by
the fraction of the 32 blocks of radius 2 which yield 1 when $f^2$ is
applied.
In general, we can write
\begin{equation}c_t = \frac{\card [\f^{-t}(1)] }{2^{2t+1}},
\end{equation} where $\card[f^{-t}(1)]$, as usual, denotes the number of preimages of 1
under $\f^t$. To make use of the table in the Appendix,
 we can rewrite the last equation as
\begin{equation}c_t= 2^{-2t-1}\sum_{\f(\beta_i)=1} \card [\f^{-t+1}(\beta_i)],
\end{equation} where the sum runs over all radius-1 blocks $\beta_i$ which yield 1
according to the cellular automaton rule, what can be also written as
\begin{equation}c_t= 2^{-2t-1}\sum_{i=0}^7 f(\beta_i) \card [\f^{-t+1}(\beta_i)].
\end{equation} Applying this procedure to rules listed in the Appendix, we obtain expressions
for $c_t$, as shown in Table \ref{density}.
\begin{table*}
\caption{Density of ones for disordered initial state with
 $c_0=0.5$.}
\label{density}
\begin{tabular}{||c|l||c|l||} \hline
Rule & $c_t$ & Rule & $c_t$ \\ \hline $1$ & $ 7/16 +
\frac{5}{16}(-1)^t$ & $2$ & $ 1/8$ \\ \hline $3$ & $ 7/16 +
\frac{3}{16}(-1)^t$ & $4$ & $ 1/8$ \\ \hline $5$ & $ 7/16 +
\frac{3}{16}(-1)^t$ & $7$ & $ \frac{15}{32}+\frac{3}{32}(-1)^t
-(-2)^{-t-4} -2^{-t-4}$ \\ \hline $8$ & $ 0$ & $10$ & $ 1/4$ \\
\hline $12$ & $ 1/4$ & $13$ & $ 7/16 - (-2)^{-t-3}$ \\ \hline $19$
& $ 1/2   + \frac{3}{32}(-1)^t$ & $23$ & $ 1/2$ \\ \hline $24$ & $
3/16$ & $27$ & $ 17/32 + \frac{1}{32}(-1)^t$ \\ \hline $28$ & $
\frac{1}{2} + \frac{1}{48}(-1)^t - \frac{5}{24}2^{-t}$ & $29$ & $
1/2$ \\ \hline $32$ & $ 2^{-1 - 2t}$ & $34$ & $ 1/4$ \\ \hline
$36$ & $ 1/16$ & $38$ & $ 9/32$ \\ \hline $40$ & $ 2^{-t-1}$ &
$42$ & $ 3/8$ \\ \hline $44$ & $ 1/6 + \frac{5}{6}2^{-2t}$ & $46$
& $ 3/8$ \\ \hline $50$ & $ 1/2 - 2^{-2t-1}$ & $72$ & $ 1/8$ \\
\hline $76$ & $ 3/8$ & $77$ & $ 1/2$ \\ \hline $78$ & $ 9/16$ &
$108$ & $ 5/16$ \\ \hline $128$ & $ 2^{-1 - 2t}$ & $130$ & $ 1/6 +
\frac{1}{3}2^{-2t}$ \\ \hline $132$ & $ 1/6 + \frac{1}{3}2^{-2t}$
& $136$ & $ 2^{-t-1}$ \\ \hline $138$ & $ 3/8$ & $140$ & $ 1/4 +
2^{-t-2}$ \\ \hline $156$ & $ 1/2$ & $160$ & $ 2^{-t-1}$ \\ \hline
$162$ & $ 1/3 + \frac{1}{6}4^{-t}$ & $164$ & $ 1/12 -
\frac{1}{3}4^{-t} +\frac{3}{4}2^{-t} $ \\ \hline $168$ & $ 3^t
2^{-2t-1}$ & $172$ & $ \frac{1}{8} +
\frac{(10-4\sqrt{5})(1-\sqrt{5})^t +
                 (10+4\sqrt{5})(1+\sqrt{5})^t}{40 \cdot 2^{2t}} $ \\ \hline
$178$ & $ 1/2$ & $200$ & $ 3/8$ \\ \hline $232$ & $ 1/2$ &  & \\
\hline
\end{tabular}
\end{table*}
 Three kinds of $c_t$ behavior can
be observed in this table:
\begin{enumerate}
\item $c_t$ is constant, like in rule 4,
\item $c_t$ oscillates and the asymptotic density is undefined, like
   in rule 5,
\item $c_t$ converges exponentially to the final density like in rule 44,
 sometimes oscillating like in rule 13.
\end{enumerate}
Note that no rule listed in Table \ref{density} converges to the
final density slower than exponentially. This is due to the fact
that all rules for which we were able to conjecture exact
expressions for the number of $n$-step preimages were either class
1 or class 2 rules according to Wolfram's classification. It is
well known that some class 3 and class 4 rules (e.g. rule 18)
exhibit power law relaxation to the final state, but we failed to
find any patterns in their n-step preimage sequences, thus no
expressions for $c_t$ could be postulated.

However, even for ``simple'' rules like those listed in Table
\ref{density}, our method yields some interesting results. For
example, \cite{Wolfram94} lists asymptotic densities for all
``minimal'' elementary rules, but for many of them only
experimental (i.e. computer simulation) values are given. For
eight such rules  we were able to find exact values of $c_\infty$,
simply by computing the limit of $c_t$ as $t \rightarrow \infty$.
These rules are presented in Table \ref{exact}, along with
experimental values of $c_\infty$ quoted after \cite{Wolfram94}.
\begin{table}
\begin{center}
\begin{tabular}{||c|l|l||} \hline
Rule    &Approximate $c_\infty$ & Exact $c_\infty$ \\ \hline
7       & $0.469 \pm 0.001$ & $15/32$ \\ \hline
13      & $0.437 \pm 0.001$ & $7/16$ \\ \hline
27      & $0.531 \pm 0.001$ & $17/32$ \\ \hline
44      & $0.167 \pm 0.001$ & $1/6$ \\ \hline
78      & $0.562 \pm 0.001$ & $9/16$ \\ \hline
130     & $0.167 \pm 0.001$ & $1/6$ \\ \hline
162     & $0.333 \pm 0.001$ & $1/3$ \\ \hline
164     & $0.083 \pm 0.001$ & $1/12$ \\ \hline
\end{tabular}
\end{center}
\caption{Rules for which exact values of asymptotic density can be
found using $n$-step preimage counting.}
\label{exact}
\end{table}
We also verified some exact values of $c_\infty$ given in \cite{Wolfram94}.
For example, the density of nonzero sites for rule 132 is
\begin{equation}c_t=\frac{1}{6}+\frac{2^{-2t}}{3},
\end{equation} hence $c_\infty=\frac{1}{6}$, not $\frac{1}{8}$ as \cite{Wolfram94} suggests.
\section{Conclusion and Remarks}
We presented some experimental results regarding sequences of numbers of
$n$-step preimages  under elementary cellular
automata rules. Many of such sequences exhibit apparent regularities,
and the  expressions for the general term of the sequence can be conjectured
for 46 out of 88 ``minimal'' cellular automata rules. Expressions
obtained this way can be used to find asymptotic emulators of rules as well as
the density of nonzero sites.

All rules discussed in this paper were either class 1 or class 2
according to Wolfram classification. Sequences of preimage numbers
for chaotic rules (except surjective rules) appear to be much more
complex, and no patterns seem to appear. If any regularities
exist, their detection will most certainly require computation of
many more terms of the sequence, and a more efficient algorithm
may be necessary. P. Grassberger \cite{Grassberger87} proposed
such an algorithm, but even with his method going beyond $n=20$
becomes unpractical. Another method proposed in \cite{Gutowitz96},
called the statistical inverse iteration, is unfortunately only
approximate, thus not very usable for the purpose of exact
enumeration.
\section{Acknowledgements}
The author acknowledges financial support from the Natural Sciences and
Engineering Research Council of Canada.

\appendix
\section{Table of preimage sequences}
The table below shows the sequences numbers of $n$-step preimages for some
elementary cellular automata rules.
They are presented as a set of 8 expressions (although not all of them are
independent), each
representing $a_n=\card [\f^n(\beta_i)]$ for all 8 basic blocks
$i=1 \ldots 7$. Only rules for which the author was able to conjecture
all 8 expressions are shown. $I_n=1$ when $n=1$, otherwise $I_n=0$.
\begin{eqnarray*}
  \mathrm{Rule\  } 0& : & 32\cdot {4^{n - 1}},0,0,0,0,0,0,0\\
  \mathrm{Rule\  } 1& : & {{-5\cdot {{\left( -4 \right) }^n} + 7\cdot {4^n}}\over 2},2\cdot {4^{n - 1}},
   {{-3\cdot {{\left( -4 \right) }^n} + 5\cdot {4^n}}\over {32}},
   {{3\cdot {{\left( -4 \right) }^n} + 11\cdot {4^n}}\over {32}},\\
& &2\cdot {4^{n - 1}},0,
   {{3\cdot {{\left( -4 \right) }^n} + 11\cdot {4^n}}\over {32}},
   {{77\cdot {{\left( -4 \right) }^n} + 85\cdot {4^n}}\over {32}}\\
  \mathrm{Rule\  } 2& : & 20\cdot {4^{n - 1}},{4^n},{4^n},0,{4^n},0,0,
   0\\
\mathrm{Rule\  } 3& : & {{-3\cdot {{\left( -4 \right) }^n} + 5\cdot {4^n}}\over 2},
   {4^n},{{-{{\left( -4 \right) }^n} + 3\cdot {4^n}}\over 8},
   {{{{\left( -4 \right) }^n} + 5\cdot {4^n}}\over 8},{4^n},\\
& &0,
   {{{{\left( -4 \right) }^n} + 5\cdot {4^n}}\over 8},
   {{11\cdot {{\left( -4 \right) }^n} + 15\cdot {4^n}}\over 8}\\
  \mathrm{Rule\  } 4& : & 21\cdot {4^{n - 1}},3\cdot {4^{n - 1}},{4^n},0,3\cdot {4^{n - 1}},
   {4^{n - 1}},0,0\\
  \mathrm{Rule\  } 5& : & {{-6\cdot {{\left( -4 \right) }^n} + 9\cdot {4^n}}\over
     4},{{-3\cdot {{\left( -4 \right) }^n} + 9\cdot {4^n}}\over {16}},5\cdot {4^{n - 1}},
   {{3\cdot {{\left( -4 \right) }^n} + 7\cdot {4^n}}\over {16}},\\
& &{{-3\cdot {{\left( -4 \right) }^n} + 9\cdot {4^n}}\over {16}},
   {{3\cdot {{\left( -4 \right) }^n} + 9\cdot {4^n}}\over 8},
   {{3\cdot {{\left( -4 \right) }^n} + 7\cdot {4^n}}\over {16}},
   {{9\cdot {{\left( -4 \right) }^n} + 11\cdot {4^n}}\over 8}\\
  \mathrm{Rule\  } 7& : & {{-3\cdot {{\left( -4 \right) }^n}}\over 4} +
    {{3\cdot {{\left( -2 \right) }^n}}\over 8} - {{17\cdot {2^n}}\over 8} +
    {{13\cdot {4^n}}\over 4},{{-{{\left( -2 \right) }^n}}\over 8} +
    {{3\cdot {2^n}}\over 8} + {{{4^n}}\over 2},\\
& &3\cdot {2^{n - 1}},
   {{{{\left( -2 \right) }^n}}\over 4} + {{3\cdot {2^n}}\over 4} +
    {{{4^n}}\over 2},{{-{{\left( -2 \right) }^n}}\over 8} +
    {{3\cdot {2^n}}\over 8} + {{{4^n}}\over 2},\\
& &{{3\cdot \left( {{\left( -2 \right) }^n} + 5\cdot {2^n} \right) }\over 8},
   {{{{\left( -2 \right) }^n}}\over 4} + {{3\cdot {2^n}}\over 4} +
    {{{4^n}}\over 2},{{3\cdot {{\left( -4 \right) }^n}}\over 4} -
    {{\left( -2 \right) }^n} - {{7\cdot {2^n}}\over 2} + {{11\cdot {4^n}}\over 4}\\
  \mathrm{Rule\  } 8& : & 8 \cdot 4^n-12 \I,4 \I,4 \I, 0, 4 \I,0,0,0\\
  \mathrm{Rule\  } 10& : & 12\cdot {4^{n - 1}},6\cdot {4^{n - 1}},{4^n},2\cdot {4^{n - 1}},
   6\cdot {4^{n - 1}},0,2\cdot {4^{n - 1}},0\\
  \mathrm{Rule\  } 12& : & 10\cdot {4^{n - 1}},6\cdot {4^{n - 1}},8\cdot {4^{n - 1}},0,6\cdot {4^{n - 1}},
   2\cdot {4^{n - 1}},0,0\\
  \mathrm{Rule\  } 13& : &
   {{-3\cdot {{\left( -2 \right) }^n}}\over 4} + {{7\cdot {2^n}}\over 4},
   {{-{{\left( -2 \right) }^n}}\over 2} + {4^n},
   {{-{{\left( -2 \right) }^n}}\over 8} - {{21\cdot {2^n}}\over 8} +
    {{7\cdot {4^n}}\over 2},\\
& &{{3\cdot {{\left( -2 \right) }^n}}\over 8} +
    {{7\cdot {2^n}}\over 8},{{-{{\left( -2 \right) }^n}}\over 2} + {4^n},
   {{3\cdot {{\left( -2 \right) }^n}}\over 4} - {{7\cdot {2^n}}\over 4} +
    {{5\cdot {4^n}}\over 2},\\
& &{{3\cdot {{\left( -2 \right) }^n}}\over 8} +
    {{7\cdot {2^n}}\over 8},{{3\cdot {{\left( -2 \right) }^n}}\over 8} +
    {{7\cdot {2^n}}\over 8}\\
   \mathrm{Rule\  } 19& : & 3 \cdot (-4)^{n-1} +5\cdot 2^{2n-1},3\cdot 4^{n-1},
   0,3\cdot 4^{n-1} +\I, 3\cdot 4^{n-1}, \I, 3\cdot 4^{n-1} +\I, \\
& &   5\cdot 2^{2n-1} - 3\cdot (-4)^{n-1} -3\I \\
   \mathrm{Rule\  } 23& : & -2 + {{1 + 2\cdot {4^{n + 1}}}\over 3},
   {{1 + 2\cdot {4^n}}\over 3},1,{{1 + 2\cdot {4^n}}\over 3},{{1 + 2\cdot {4^n}}\over 3},
   1,{{1 + 2\cdot {4^n}}\over 3},\\
& &-2 + {{1 + 2\cdot {4^{n + 1}}}\over 3}\\
\mathrm{Rule\  } 24& : &14\cdot 4^{n-1} -4\I,6\cdot 4^{n-1},6\cdot 4^{n-1} +2\I,
 0,6\cdot 4^{n-1} ,2\I,0,0 \\
  \mathrm{Rule\  } 27& : & (-4)^{n-1} + 5\cdot4^{n-1} ,
   {4^n},2\cdot {4^{n - 1}},{4^n},{4^n},
   2\cdot {4^{n - 1}},{4^n},-(-4)^{n-1}+7 \cdot 4^{n-1}\\
  \mathrm{Rule\  } 28& : & {{5\cdot {2^n}}\over 2},{{-{{\left( -4 \right) }^n}}\over 6} +
    {{{{\left( -2 \right) }^n}}\over 2} + {{5\cdot {2^n}}\over 3} +
    {{{4^n}}\over 2},{{-{{\left( -4 \right) }^n}}\over 6} -
    {{\left( -2 \right) }^n} - {{10\cdot {2^n}}\over 3} + 3\cdot {4^n},\\
& &{{{{\left( -4 \right) }^n}}\over 6} +
    {{{{\left( -2 \right) }^n}}\over 2} + {{5\cdot {2^n}}\over 6} +
    {{{4^n}}\over 2},{{-{{\left( -4 \right) }^n}}\over 6} +
    {{{{\left( -2 \right) }^n}}\over 2} + {{5\cdot {2^n}}\over 3} +
    {{{4^n}}\over 2},\\
& &{{{{\left( -4 \right) }^n}}\over 6} -
    {{\left( -2 \right) }^n} - {{25\cdot {2^n}}\over 6} + 3\cdot {4^n},
   {{{{\left( -4 \right) }^n}}\over 6} +
    {{{{\left( -2 \right) }^n}}\over 2} + {{5\cdot {2^n}}\over 6} +
    {{{4^n}}\over 2},0\\
   \mathrm{Rule\  } 29& : & 3\cdot {4^{n - 1}},3\cdot {4^{n - 1}},
   7\cdot {4^{n - 1}},3\cdot {4^{n - 1}},3\cdot {4^{n - 1}},7\cdot {4^{n - 1}},
   3\cdot {4^{n - 1}},3\cdot {4^{n - 1}}\\
  \mathrm{Rule\  } 32& : & -11 + 32\cdot {4^{n - 1}},3,4,0,3,1,0,0\\
  \mathrm{Rule\  } 34& : & 10\cdot {4^{n - 1}},6\cdot {4^{n - 1}},8\cdot {4^{n - 1}},0,6\cdot {4^{n - 1}},
   2\cdot {4^{n - 1}},0,0\\
\mathrm{Rule\  } 36& : &26\cdot 4^{n-1} -10\I,2\cdot 4^{n-1}+2\I,2\cdot 4^{n-1},
2\I,2\cdot 4^{n-1}+2\I,0,2\I,2\I \\
\mathrm{Rule\  } 38& : & 47 \cdot 4^{n-2} -\frac{11}{4}\I, 21 \cdot 4^{n-2}-\frac{1}{4}\I,
3\cdot 4^{n-1}, 3\cdot 4^{n-1}, 21 \cdot 4^{n-2} - \frac{1}{4}\I,\\
& &3\cdot 4^{n-2}+ \frac{1}{4}\I, 3\cdot 4^{n-1}, 3\I \\
  \mathrm{Rule\  } 40& : & -9\cdot {2^n} + 32\cdot {4^{n - 1}},4\cdot {2^{n - 1}},
   4\cdot {2^{n - 1}},{2^n},4\cdot {2^{n - 1}},{2^n},
   {2^n},0\\
   \mathrm{Rule\  } 42& : & 7\cdot {4^{n - 1}},5\cdot {4^{n - 1}},{4^n},
   {4^n},5\cdot {4^{n - 1}},3\cdot {4^{n - 1}},{4^n},0\\
\mathrm{Rule\  } 44& : & -7+4^{n+1}, \frac{1}{3}(4^{n+1}-1),
\frac{1}{3}(4^{n+1}-1) -3 +\I, 4-\I, \frac{1}{3}(4^{n+1}-1), \\
& &1,4-\I,2+\I \\
\mathrm{Rule\  } 46& : & 38 \cdot 4^{n-2} -7\cdot \I/2,18\cdot 4^{n-2}-\I/2,
0,6\cdot 4^{n-1},\\
& &18\cdot 4^{n-2} -\I/2, 6\cdot 4^{n-2} +\I/2, 6 \cdot 4^{n-1}, 4\I \\
  \mathrm{Rule\  } 50& : & 5,3 + {{2\cdot \left( -1 + {4^n} \right) }\over 3},
   {{8\cdot \left( -1 + {4^n} \right) }\over 3},
   {{2\cdot \left( -1 + {4^n} \right) }\over 3},
   3 + {{2\cdot \left( -1 + {4^n} \right) }\over 3},\\
& &-3 + {{8\cdot \left( -1 + {4^n} \right) }\over 3},
   {{2\cdot \left( -1 + {4^n} \right) }\over 3},0\\
\mathrm{Rule\  } 72& : & 97\cdot 4^{n-2} -41\I/4,7\cdot 4^{n-2} +9\I/4,
4\I, 2^{2n-1},7\cdot 4^{n-2} +9\I/4,\\
& & 4^{n-2} +7\I/4, 2^{2n-1},0 \\
  \mathrm{Rule\  } 76& : & 5\cdot {4^{n - 1}},5\cdot {4^{n - 1}},8\cdot {4^{n - 1}},2\cdot {4^{n - 1}},
   5\cdot {4^{n - 1}},5\cdot {4^{n - 1}},2\cdot {4^{n - 1}},0\\
  \mathrm{Rule\  } 77& : & 1,{{1 + 2\cdot {4^n}}\over 3},-2 + {{1 + 2\cdot {4^{n + 1}}}\over 3},
   {{1 + 2\cdot {4^n}}\over 3},{{1 + 2\cdot {4^n}}\over 3},\\
& &-2 + {{1 + 2\cdot {4^{n + 1}}}\over 3},{{1 + 2\cdot {4^n}}\over 3},1\\
 \mathrm{Rule\  } 78& : &3\cdot 2^{n-1},3\cdot 2^{n-1},5\cdot 2^{2n-1}-9\cdot 2^{n-1}+4\I,
3\cdot 2^{n-1}+4^n-2\I, 3\cdot 2^{n-1},\\
& &7\cdot 2^{2n-1}-9\cdot 2^{n-1}+2\I,
3\cdot 2^{n-1}+4^n-2\I,3\cdot 2^{n-1}-2\I \\
 \mathrm{Rule\  } 108& : & 32\cdot 4^{n-2}-3\I,26\cdot 4^{n-2}-3\I/2,
 24\cdot 4^{n-2}-3\I,6\cdot 4^{n-2}+3\I/2,\\
 & &26\cdot4^{n-2}-3\I/2,4^{n-1},6\cdot 4^{n-2}+3\I/2,4^{n-1} +6\I \\
  \mathrm{Rule\  } 128& : & -8 + 32\cdot {4^{n - 1}},2,1,1,2,0,1,1\\
  \mathrm{Rule\  } 130& : & -3 + {4^{n + 1}},{{-1 + {4^{n + 1}}}\over 3},
   {{-1 + {4^{n + 1}}}\over 3},1,{{-1 + {4^{n + 1}}}\over 3},1,1,1\\
  \mathrm{Rule\  } 132& : & -4 + {{17\cdot {4^n}}\over 4},{2\over 3} + {{13\cdot {4^n}}\over {12}},
   {{-1 + {4^{n + 1}}}\over 3},1,{2\over 3} + {{13\cdot {4^n}}\over {12}},
   {4^{n - 1}},1,1\\
  \mathrm{Rule\  } 136& : & -8\cdot {2^n} + 8\cdot {4^n},4\cdot {2^{n - 1}},
   {2^n},{2^n},4\cdot {2^{n - 1}},0,{2^n},
   {2^n}\\
   \mathrm{Rule\  } 138& : & 8\cdot {4^{n - 1}},6\cdot {4^{n - 1}},3\cdot {4^{n - 1}},
   3\cdot {4^{n - 1}},6\cdot {4^{n - 1}},0,3\cdot {4^{n - 1}},3\cdot {4^{n - 1}}\\
  \mathrm{Rule\  } 140& : & -2\cdot {2^n} + {{5\cdot {4^n}}\over 2},6\cdot {4^{n - 1}},-{2^n} + 2\cdot {4^n},
   {2^n},6\cdot {4^{n - 1}},2\cdot {4^{n - 1}},\\
& &{2^n},
   {2^n}\\
   \mathrm{Rule\  } 156& : & {2^n},{2^n} + {{{4^n}}\over 2},
   -3\cdot {2^n} + 3\cdot {4^n},{2^n} + {{{4^n}}\over 2},{2^n} + {{{4^n}}\over 2},
   -3\cdot {2^n} + 3\cdot {4^n},\\
& &{2^n} + {{{4^n}}\over 2},{2^n}\\
  \mathrm{Rule\  } 160& : & 3 - 10\cdot {2^n} + 8\cdot {4^n},-1 + {2^{n + 1}},-3 + 4\cdot {2^n},1,
   -1 + {2^{n + 1}},-1 + {2^{n + 1}},1,1\\
  \mathrm{Rule\  } 162& : & {{-1 + {4^{n + 1}}}\over 3},{{-1 + {4^{n + 1}}}\over 3},
   -2 + {{1 + 2\cdot {4^{n + 1}}}\over 3},1,{{-1 + {4^{n + 1}}}\over 3},
   {{-1 + {4^{n + 1}}}\over 3},1,1\\
  \mathrm{Rule\  } 164& : & 4 - 9\cdot {2^n} + 6\cdot {4^n},-{2\over 3} + {2^n} + {{2\cdot {4^n}}\over 3},
   {{1 + 2\cdot {4^n}}\over 3},-1 + {2^{n + 1}},\\
& &-{2\over 3} + {2^n} + {{2\cdot {4^n}}\over 3},{2^n},-1 + {2^{n + 1}},
   -1 + {2^{n + 1}}\\
  \mathrm{Rule\  } 168& : & -7\cdot {3^n} + 8\cdot {4^n},{3^n},{3^n},{3^n},{3^n},
   {3^n},{3^n},{3^n}\\
  \mathrm{Rule\  } 172& : & 20\cdot {4^{n - 1}} - {{-{{\left( 1 - {\sqrt{5}} \right) }^{n + 2}} +
        {{\left( 1 + {\sqrt{5}} \right) }^{n + 2}}}\over {8\cdot {\sqrt{5}}}} -
    {{-{{\left( 1 - {\sqrt{5}} \right) }^{n + 3}} +
        {{\left( 1 + {\sqrt{5}} \right) }^{n + 3}}}\over {4\cdot {\sqrt{5}}}},\\
& &{4^n},{4^n} -
    {{-{{\left( 1 - {\sqrt{5}} \right) }^{n + 1}} +
        {{\left( 1 + {\sqrt{5}} \right) }^{n + 1}}}\over {2\cdot {\sqrt{5}}}},\\
& &{{-\left( -{{\left( 1 - {\sqrt{5}} \right) }^{n + 2}} +
          {{\left( 1 + {\sqrt{5}} \right) }^{n + 2}} \right) }\over
      {8\cdot {\sqrt{5}}}} + {{-{{\left( 1 - {\sqrt{5}} \right) }^{n + 3}} +
        {{\left( 1 + {\sqrt{5}} \right) }^{n + 3}}}\over {8\cdot {\sqrt{5}}}},\\
& &{4^n},{{-{{\left( 1 - {\sqrt{5}} \right) }^{n + 2}} +
       {{\left( 1 + {\sqrt{5}} \right) }^{n + 2}}}\over {8\cdot {\sqrt{5}}}},\\
& &{{-\left( -{{\left( 1 - {\sqrt{5}} \right) }^{n + 2}} +
          {{\left( 1 + {\sqrt{5}} \right) }^{n + 2}} \right) }\over
      {8\cdot {\sqrt{5}}}} + {{-{{\left( 1 - {\sqrt{5}} \right) }^{n + 3}} +
        {{\left( 1 + {\sqrt{5}} \right) }^{n + 3}}}\over {8\cdot {\sqrt{5}}}},\\
& &{{-{{\left( 1 - {\sqrt{5}} \right) }^{n + 3}} +
       {{\left( 1 + {\sqrt{5}} \right) }^{n + 3}}}\over {8\cdot {\sqrt{5}}}}\\
  \mathrm{Rule\  } 178& : & 1,{{1 + 2\cdot {4^n}}\over 3},-2 + {{1 + 2\cdot {4^{n + 1}}}\over 3},
   {{1 + 2\cdot {4^n}}\over 3},{{1 + 2\cdot {4^n}}\over 3},\\
& &-2 + {{1 + 2\cdot {4^{n + 1}}}\over 3},{{1 + 2\cdot {4^n}}\over 3},1\\
  \mathrm{Rule\  } 200& : & 13\cdot {4^{n - 1}},3\cdot {4^{n - 1}},0,{4^n},3\cdot {4^{n - 1}},
   {4^{n - 1}},{4^n},{4^n}\\
  \mathrm{Rule\  } 232& : & -2 + {{1 + 2\cdot {4^{n + 1}}}\over 3},{{1 + 2\cdot {4^n}}\over 3},1,
   {{1 + 2\cdot {4^n}}\over 3},{{1 + 2\cdot {4^n}}\over 3},1,{{1 + 2\cdot {4^n}}\over 3},\\
& &-2 + {{1 + 2\cdot {4^{n + 1}}}\over 3} \\
\end{eqnarray*}


\end{document}